\documentclass[pra,aps,twocolumn,showpacs]{revtex4}
\input{epsf}
\begin{document}
\title{Finite size effects in the dynamics and thermodynamics of 2D Coulomb
clusters}
\author{E. Yurtsever}
\affiliation{Ko\c c University, Rumelifeneriyolu, Sariyer, Istanbul 34450,
Turkey}
\author{F. Calvo}
\affiliation{Laboratoire de Physique Quantique, IRSAMC, Universit\'e Paul
Sabatier, 118 Route de Narbonne, F31062 Toulouse Cedex, France}
\author{D. J. Wales}
\affiliation{University Chemical Laboratories, Cambridge CB2 1EW,
United Kingdom}
\begin{abstract}
The dynamics and thermodynamics
of melting in two-dimensional Coulomb clusters is revisited
using molecular dynamics and Monte Carlo simulations. Several
parameters are considered, including the Lindemann index, the largest
Lyapunov exponent and the diffusion constant. In addition to the
orientational and radial melting processes, isomerizations and complex
size effects are seen to occur in a very similar way to atomic and molecular
clusters. The results are discussed in terms of the
energy landscape represented through disconnectivity
graphs, with proper attention paid to the broken ergodicity
problems in simulations. Clusters bound by $1/r^3$ and
$e^{-\kappa r}/r$ forces,
and heterogeneous clusters made of singly- and doubly-charged species,
are also studied, as well as the evolution toward larger systems.
\end{abstract}
\maketitle

\section{Introduction}
\label{sec:intro}

The finite size analogs of two-dimensional (2D) Wigner crystals have received 
considerable attention from theoreticians
\cite{lozovik90,lozovik92,bedanov94,schweigert95,peeters95b,partoens97,schweigert98b,candido98,schweigert98,juan98,koulakov98,lozovik98,lai99,schweigert00,kong02,drocco03},
and more recently also from experimental groups
\cite{melzer96,bubeck99,stjean01,melzer03,ichiki04}.
Particles in mutual repulsion can be
confined using a variety of methods, most importantly electrostatic traps of
the Paul or Penning types, and simple hard walls, usually of circular
shape. The repulsion between particles depends upon the details of the experiment.
Arrays of electrons on the surface of liquid helium
\cite{leiderer92} or in quantum dots \cite{dots} can be described
by conventional Coulomb
forces. However, logarithmic forces are more relevant for vortex clusters in
superfluids \cite{yarmachuk82} or Bose-Einstein condensates \cite{chevy00}.
Mesoscopic colloidal particles undergo dipolar $1/r^3$ interactions within
an external magnetic field \cite{zahn97}, or a Yukawa (screened Coulomb)
potential when they are charged and placed in a viscous medium. 

Finite 2D assemblies have also been synthesised on a more
macroscopic scale, using metal particles close to 1\,mm in size
\cite{stjean01}. In this study, the authors were able to visualise and
locate not only stable configurations, but they could also estimate
the shape of the transition states, which are here defined as
stationary points with Hessian index one \cite{walesbook}.

Another example of experimentally accessible Wigner crystals occurs for
dusty plasmas. These highly charged particles of micrometer size
and charges up to $Z=10^4$ can be stabilised as gaseous plasmas. The 
gravitational forces and the electric field balance each other, so that 
the dust particles form highly regular two-dimensional structures. By changing the 
electric field, or conversely the pressure, of the plasma, transitions from
solid to liquid-like phases are observed \cite{melzer96}. A two-stage
melting was proposed, which does not follow the predictions of two-dimensional
melting theories, but rather goes through an intermediate phase. This phase
has been described as an oscillating crystal \cite{ichiki04}. Finally,
by measuring the spectra of dust particles, it has been suggested
that the dynamics may be a size-dependent phenomenon, that is the
effects of various types of motion such as intershell rotation and
breathing may play different roles \cite{melzer03}.
 
A major part of our theoretical understanding of the above systems comes from
numerical studies.
The static properties of the assemblies were obtained using optimization
methods ranging from simulated annealing to genetic algorithms, and a 
Mendeleev-type table was proposed by several authors
\cite{lozovik90,peeters95b} to reflect the special stability
of certain sizes. Comparisons with the
theoretical predictions of the `Thomson atom' \cite{thomson} were also
carried out \cite{partoens97}. Shell effects were more generally investigated
semi-analytically by Koulakov and Shklovskii \cite{koulakov98}. Metastable
configurations (excited states) and the potential energy barriers between them have
been studied in Refs.~\cite{schweigert95}, \cite{lozovik98},
\cite{kong02} and \cite{wales93}. 
Topological defects were also identified for stable structures \cite{lai99}.

Melting in 2D Coulomb clusters has been a subject of debate. The results of
the classical and quantum Monte Carlo simulations of Lozovik
and Mandelshtam \cite{lozovik92} were interpreted as a two-step
melting process, where orientational melting occurs prior to radial melting.
In this process, shells remain quasi-crystalline, but still undergo some
concerted rotational motion. Radial melting conversely involves exchanges
between particles of adjacent shells. This two-step process is generally
accepted for small assemblies, but it is not clear whether, in large clusters,
the external shell may also exhibit orientational melting \cite{lozovik98} or
not \cite{schweigert98}.

Several factors are likely to influence the stability of 2D Coulomb clusters.
First, rotational barriers are strongly dependent on the structure,
which in turn depends on the size. More stable structures are
obtained when the core is based on a triangular lattice, thus resembling most
observed Wigner crystals. More importantly, the very different nature of
orientational and radial melting has led to some confusion about how 
suitable observables should be defined to probe these processes separately
\cite{schweigert00,rinn01,schweigert01}. In their study of binary 2D clusters,
Drocco {\em et al.}~\cite{drocco03} chose an order parameter based on the polar
angles of the particles in order to quantify the extent of orientational
motion in Langevin molecular dynamics simulations. However, this measure becomes
impractical as soon as a central particle is present, as it then becomes
very noisy.

Melting in atomic and molecular clusters also displays a very rich
phenomenology \cite{calvorev}. In particular, surface melting \cite{cheng92},
plastic transitions \cite{fuchs98} or anomalously high melting temperatures
\cite{jarrold03} have been identified from experiments or simulations. 
For three-dimensional trapped-ion clusters, the structures
are sometimes related to those found in atomic clusters \cite{wales93}.
The study of atomic clusters also shares with Coulomb assemblies
the lack of a proper universal order parameter to characterise the melting
point. For instance, the Lindemann criterion for melting does not always
provide useful information in the case of solid-solid transitions and
premelting effects \cite{cs2000}.

The largest Lyapunov exponent $\lambda$, which quantifies chaos and
the sensitivity to initial conditions, is a 
useful tool for probing the complex dynamics of atomic and molecular
clusters \cite{hinde92,amitrano92,yurtsever97,calvo98,calvomolec}, especially
at melting. Because Lyapunov exponents are related to the second
derivatives of the potential energy surface (PES) \cite{wales91},
we expect them to be sensitive to orientational melting.

In this paper, we revisit the dynamics and thermodynamics of 2D Coulomb
clusters by focusing on dynamical measures such as $\lambda$ or the
diffusion coefficient, $D$, and by performing extensive Monte
Carlo simulations employing the parallel tempering strategy \cite{ptmc}.
We also consider the interplay between the dynamics and the energy landscape
of these clusters by constructing disconnectivity graphs \cite{becker97,walesmw98}.
Finally, we investigate how the thermodynamic behaviour evolves with size.

This article is organised as follows. After briefly presenting our methods,
we choose some specific examples to illustrate the difficulties associated
with an unambiguous definition of the melting point in these clusters, and
we then discuss the large size variations. We finally summarise and conclude
in Section IV. 

\section{Methods}
\label{sec:methods}

2D Wigner clusters are characterised by the following classical
potential energy:
\begin{equation}
E({\bf R}) = \sum_{i<j} \frac{q_iq_j}{r_{ij}^p}\exp(-\kappa r_{ij})
+ \sum_i A r_i^2.
\label{eq:energy}
\end{equation}
In this equation, $\{q_i\} =1$ or 2 are the charges carried by the
particles, $r_{ij}$ is the distance between particles $i$ and $j$, and $r_i$
is the distance from particle $i$ to the cluster centre-of-mass. The
parameter $p$ will usually be $1$ (Coulomb
case), but we also specifically studied the case $p=3$ (dipolar
interaction) for which some results will be selected below. The range
of the electrostatic potential, $\kappa$, was set to 0 (pure Coulomb)
or to a finite value (shielded Coulomb). Finally, and for comparison
with the work reported in Ref.~\onlinecite{drocco03} on binary clusters, we
have taken $A=10$. It should be noted that, in the absence of the
exponential term, the physics of the system is independent 
of $A$: under the transformation
$(r,E)\to(rA^{1/(2+p)},E/A^{p/(2+p)})$, the energy can be cast in a
form that does not depend on $A$. In the following presentation, unit masses and
charges are used and all properties are explicitly given in dimensionless reduced units. 

The global minima of the clusters were investigated using the
Monte Carlo plus minimization algorithm, also known as basin-hopping
\cite{li87,wales97}. The 2D clusters can generally be described
as multishell systems, and we will use the $(n_1,n_2...)$ notation for
a homogeneous cluster having $n_i$ particles in its $i^{\rm th}$
shell. Binary clusters will be referred to as $\{N_s,N_d\}$, where $N_s$
and $N_d$ are the respective numbers of singly- and doubly-charged particles.
Due to the relatively long range of the potential \cite{walesbook}, global optimisation
is relatively straightforward for Coulomb systems.

Beyond the stable isomers, we also investigated higher-index
stationary points on the energy landscape. Locating the 
saddle points with Hessian index one
enabled us to construct disconnectivity graphs
\cite{walesbook}, in order to relate the observed dynamical and
thermodynamical behaviour to the potential energy surface for these clusters.
We refer the reader to Ref.~\onlinecite{walesbook} for
further details of how stationary points are located and
disconnectivity graphs are generated.

The dynamics has also been studied by solving 
Hamilton's equations of motion at constant energy. The coupled 
equations were integrated by the 4-step Runge-Kutta method, which keeps 
the energy constant to at least seven digits. For the first trajectory,
the initial positions of the particles were obtained from the global 
minimum structure and a random set of momenta were chosen.
The momenta were then scaled to set the linear and angular momentum to 
zero and the vibrational temperature to around 
$10^{-7}$. The time step was taken to be 0.0025. The first $10^5$ steps 
were used to equilibrate the system and discarded, then $10^6$ steps 
were used for each simulation. Further trajectories start with the
configuration obtained from the previous run, with velocities scaled approximately to give a new 
temperature increased by a factor 1.25. Again, the equilibration steps
were discarded.

For each trajectory the average kinetic temperature, root-mean-square (RMS) bond length
fluctuation index $\delta$, the maximal Lyapunov exponent, $\lambda$, (MLE) and 
diffusion coefficient, $D$, were calculated. $\delta$ is commonly
used to detect melting in bulk or finite (3D) atomic
systems, as it is expected to jump from a low value to around
0.1--0.15. Both the diffusion coefficient and the Lyapunov exponent
were calculated from shorter parts of the entire trajectory. 
$D$ was obtained from the slope of the mean square atomic
displacement. This measure describes the breathing and inter-shell
exchanges (isomerizations), but does not identify even qualitative
changes in the rotational structure. For homogeneous
systems it is very difficult to define measures of the angular
motion. Since the total angular momentum is zero, one can only measure
angular velocity and momenta for separate shells; however, the shell
definitions lose their significance once isomerizations start
taking place. In addition, the angular properties become ill-defined
for clusters having a central particle.

The calculation of the MLE was motivated by several studies indicating
possible relations between phase changes and this dynamical
indicator. The MLE measures the sensitive dependence on initial conditions
and is defined from the exponential separation of two trajectories 
that start infinitesimally close together in the phase space. 
A tangent space approach was used in the numerical calculations \cite{benettin74}.
However, we should
mention that semi-analytical theories have also been proposed to
quantify the MLE from the statistical properties of the potential
energy surface \cite{casetti95}.

The equilibrium thermodynamics of the 2D clusters was also studied by
means of parallel tempering Monte Carlo (MC) simulations. Parallel
tempering is a particularly efficient method to accelerate convergence
in systems whose natural dynamics is affected by broken ergodicity.
Most often such problems are related to multiple funnels \cite{walesbook}
or `glassy' energy landscapes. The MC simulations were performed
in the canonical ensemble, with temperatures in the range
$10^{-5}<T<10$. About 80 runs were carried out in this range,
with the temperatures
nearly equally spaced on a logarithmic scale. $10^5$ MC cycles were used
for equilibration, followed by $10^6$ MC cycles for the actual
calculations. The MC calculations were principally used to compute
the thermal properties, such as the heat capacity, $C_v$.

In both MC and MD simulations the relation between the energy and the
temperature is quasi linear.

\section{Results}

The snapshots obtained from experiments on colloidal particles
\cite{stjean01} or dust plasmas \cite{melzer03,ichiki04}, as well as
the various molecular dynamics (MD) simulations,
exhibit three types of characteristic motion.
At low energy or temperature, or for dust plasmas under high pressures,
only small amplitude motion is observed, in which the particles
oscillate around their mean positions. As the energy increases, these
oscillations gain enough momentum to achieve complete rotations in
their respective shells. The directions of these rotations are random,
and they may change in time. However, in order to preserve the
total angular momentum, different shells must rotate in opposite
directions. Eventually, actual isomerizations between shells
may take place, as previously examined in 3D systems \cite{wales93}.
In one case, it has also been
reported that isomerizations may occur before the full rotational
motion is achieved \cite{melzer03}. We expect this observation to be due to
rather infrequent sampling of the experimental trajectories \cite{melzer03}.
It is clear that the angular and the radial motion of particles make different 
contributions to all the measures aforementioned.

\begin{figure}[htb]
\setlength{\epsfxsize}{9cm}
\centerline{\epsffile{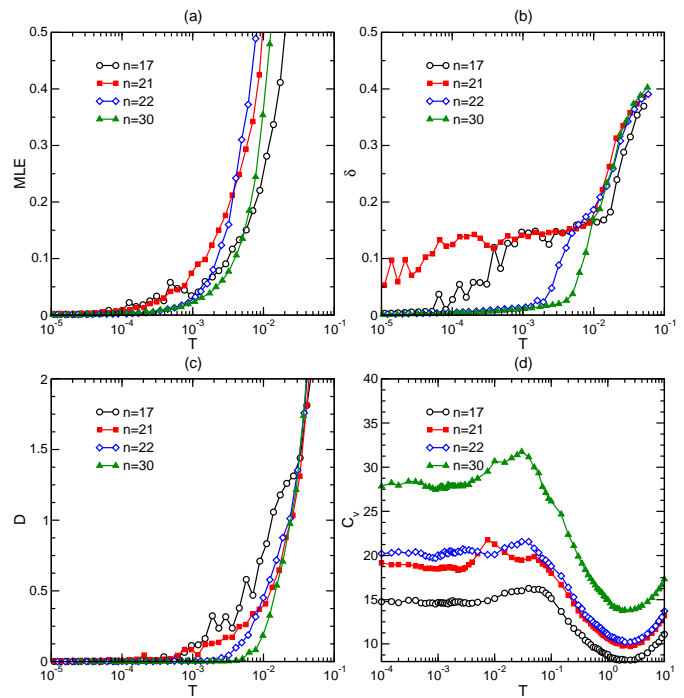}}
\caption{Dynamic and thermodynamic properties of some homogeneous
2D Coulomb clusters with $n=17$, 21, 22, and 30 as a function of temperature.
(a) Maximum Lyapunov exponent (MLE) $\lambda$; (b) RMS bond length
fluctuation $\delta$; (c) diffusion coefficient $D$; (d) canonical
heat capacity $C_v$. The data in (a) to (c) are from microcanonical
MD simulations, the data in (d) are from canonical Monte Carlo simulations.
$C_v$ is in units of the Boltzmann constant}
\label{fig:homogeneous}
\end{figure}

In order to study the mechanism of phase transitions in 2D clusters, 
we have carried out simulations of homogeneous clusters systems in the size
range $3\leq n\leq 40$. We selected several sizes displaying different
but typical characteristics. In Fig.~\ref{fig:homogeneous}, $\lambda$,
$D$, $\delta$ and $C_v$ are plotted with respect to the average
temperature for sizes 17, 21, 22 and 30. The global minima for these
clusters are found to be (1,6,10), (1,7,13), (2,8,12) and
(5,10,15), respectively, in agreement with Ref.~\onlinecite{bedanov94}.
The MLE starts becoming nonzero
at very low temperatures, around $T=10^{-4}$. In this regime, the
dynamics are more irregular for $n=17$ and 21.
Similarly, the Lindemann index and the diffusion coefficients are also
higher for $n=17$ and 21 in the low $T$ regime. Both these clusters exhibit
an intermediate phase over a broad temperature range (note the
logarithmic temperature scale).
On the other hand, the 30-particle cluster exhibits a single phase
transition.

While the three dynamical indicators agree with one another, the heat
capacities provide rather little information in comparison.
The general behaviour observed
for all sizes is that of harmonic oscillators at very low temperatures
$(C_v\to (2n-3)k_B)$ or for free, independent particles at high
temperatures ($C_v \to (2n-3)k_B$ again due to confinement). 
In the intermediate temperature range,
a strong decrease is seen near $T\sim 2$ for all sizes, following an
order-disorder peak below $T=0.1$. The curves for a specific cluster
are built on the basis of this generic behaviour, which reminds us
of the caloric curves computed for Lennard-Jones polymers
\cite{polymers}. For some sizes, such as $n=21$ or 30, extra bumps or
shoulders occur as a consequence of isomerizations. In general, the
strong increase of $\lambda$, $D$ or $\delta$ at $T\simeq 0.03$
correlates with a peak in the specific heat. However, the phenomena 
responsible for the peculiar dynamical features for the 17- and
21-particle clusters do not have any thermal signature. The premelting
heat capacity peak seen for $n=21$ at $T=0.01$ does not have any
counterpart in the dynamical indices either.

\begin{figure*}[htb]
\setlength{\epsfxsize}{14cm}
\centerline{\epsffile{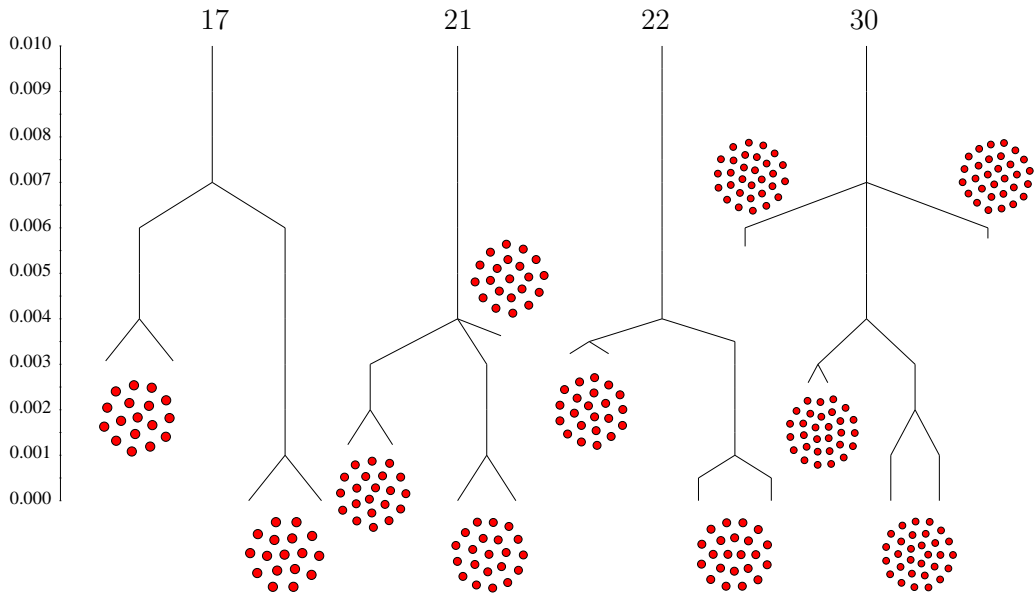}}
\caption{Disconnectivity graphs for the homogeneous clusters of
Fig.~\protect\ref{fig:homogeneous}. The global minimum energy is
shifted to zero for each size.}
\label{fig:trees}
\end{figure*}

These data can be interpreted using the disconnectivity graphs shown in
Fig.~\ref{fig:trees}. In this figure, saddles connecting two
permutations of the same isomers have been included to account for the
rotational barriers. Such saddles are manifested by the presence of
'copies' of the corresponding minima on the graph.

The energy barriers, which link the global
minimum to itself through an internal rotation, are much lower
for $n=17$ ($\Delta E=4\times 10^{-6}$) and $n=21$ ($\Delta E=10^{-9}$, this is
nearly a case of free rotation) than for larger clusters, where they are
closer to $\Delta E=10^{-3}$. Such degenerate rearrangements \cite{walesbook}
do not play a major role
in thermal equilibrium, but can have important dynamical consequences.
This is precisely what we observe in Fig.~\ref{fig:homogeneous}. The
barriers for rearrangements are much higher in energy than those
involved in the rotational motion, but in some cases other stable
isomers lie relatively close to the global minimum. For
$n=21$ and $n=30$, such isomerizations are indeed seen and
correlate with bumps in the canonical heat capacity.

\begin{figure}[htb]
\setlength{\epsfxsize}{9cm}
\centerline{\epsffile{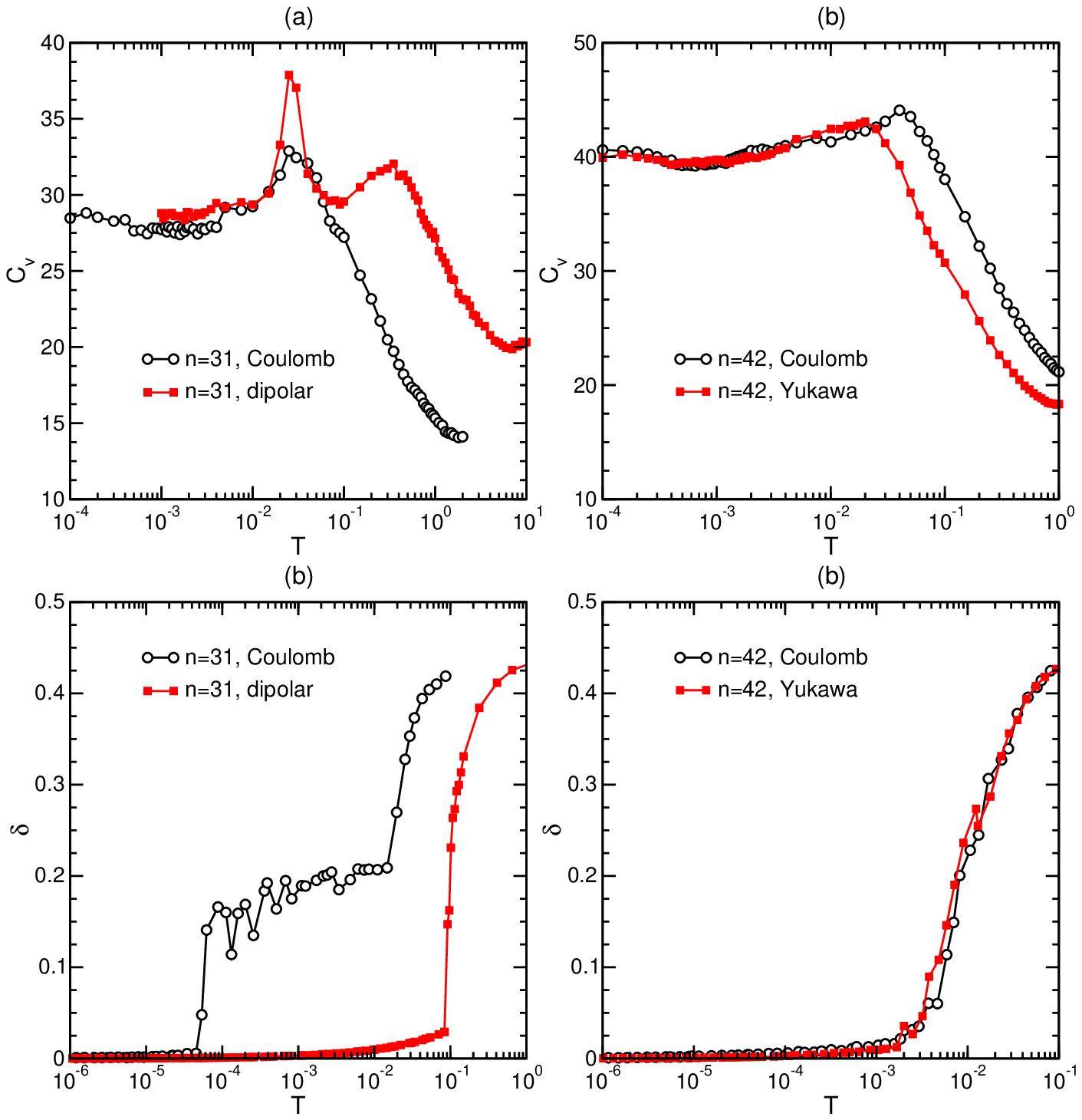}}
\caption{(a) and (b): heat capacities of 31- and 42-particle clusters,
respectively (in units of the Boltzmann constant); (c) and (d): RMS bond length fluctuation indices. For
$n=31$, the results of the Coulomb ($1/r$) and dipolar ($1/r^3$)
interactions are shown. For $n=42$, the results obtained with the
Yukawa potential ($e^{-r}/r$) are compared to those with the bare
Coulomb form.}
\label{fig:3142}
\end{figure}

>From the previous results it is clear that one cannot draw general
conclusions by looking at dynamical indicators alone, or at observables
aimed at characterising thermal equilibrium. We have also looked at
two specific cluster sizes, for which the results of the pure Coulomb
case have been compared to other interaction forms. The 31-particle
cluster was studied with both $1/r$ and $1/r^3$ interactions, the
latter expression being more relevant to colloidal particles, such as
the ones involved in the experimental setup of
Ref.~\onlinecite{stjean01}. The 42-particle cluster has recently been the
focus of an experimental report on dust plasmas \cite{ichiki04}. In this
situation, a screened Coulomb or Yukawa form $e^{-\kappa r}/r$ is more
appropriate than the bare Coulomb law. Here we chose $\kappa=1$, but
the results presented hereafter in Fig.~\ref{fig:3142} are not
significantly different for other values in the range
$0.1\leq \kappa\leq 10$. 

We first discuss the 31-particle clusters, which display different
behaviour depending on the form of the interaction. While the Coulomb
potential leads to a (5,11,15) 3-shell global minimum, the shorter range
dipolar interaction favours a (1,5,10,15) ground state. Because this
structure has a central particle, the vibrational motion is more
constrained, and the harmonic entropy of (5,11,15) is actually
larger. This competition between energy and entropy leads to a
structural transition, which is seen in the heat capacity of
Fig.~\ref{fig:3142}(a) as a significant peak near $T=0.03$, while the
melting peak is located at $T\sim 0.5$. The Coulomb
cluster behaves very differently, with a single $C_v$ peak at
$T=0.03$. Scrutiny of the Lindemann index indicates that this cluster
undergoes free internal rotation at very low temperatures $T<10^{-4}$, before
fully melting at $T=0.03$. In the case of the dipolar interaction,
only a single, but very sharp, increase is seen, at temperatures even
higher than the melting peak, $T\sim 0.1$. This surprising result has
been checked and confirmed by extensive sampling. After regular
quenching of the MD trajectories in this temperature range, we found
that the sharp increase of $\delta$ was correlated with a sudden
isomerization to the (5,11,15) second-lowest minimum, which is also involved in
the preliminary structural transition. The lowest energy transition state from
the global minimum was found to be significantly higher than
the energy of the second isomer. In such a situation, MD simulations
suffer from broken ergodicity, since they cannot sample the
available phase space until sufficient energy is added. In cluster
physics, a well-studied example of broken ergodicity is provided by the 38-atom
Lennard-Jones cluster \cite{lj38wales,lj38neirotti}. The present
sharp jump of the Lindemann index should not be interpreted as the
onset of melting, but rather as a signature of isomerization.

42-particle clusters were investigated with the aim of finding some
thermodynamic confirmation that melting proceeds in multiple steps.
Unfortunately, neither the specific heats of Fig.~\ref{fig:3142}(b) nor
the Lindemann indices of Fig.~\ref{fig:3142}(d) show any multiple-stage
behaviour that could be related to the experimental observation
\cite{ichiki04}. We thus conclude that the dust plasma system studied by
Melzer and coworkers cannot be simply modelled as if it were made of
a single layer.

Comparing the general curves for the Coulomb, dipolar and Yukawa
potentials also provides useful information about the factors
influencing the melting point.
Yukawa systems exhibit a significant shift of their melting
temperature toward lower $T$. Since the Yukawa potential is an
attenuated form of the pure Coulomb form, this is the expected trend.
At large distances, the dipolar interaction is also much smaller than
the Coulomb interaction. However, the melting temperature of Coulomb
clusters is about one order of magnitude {\em smaller}\/ than for
the dipolar cluster. Hence the long-range part of the potential plays 
opposite roles in the dipolar and Yukawa forms. This result 
suggests that the location of the melting peak is mainly driven by the
short-range repulsive part of the potential.

We now turn to small, binary Coulomb clusters, such as those
recently investigated by Drocco and coworkers \cite{drocco03}. $N_s$
particles with charge 1 and $N_d$ particles with charge 2 are
stabilised in a parabolic trap with $A=10$.
The structures again consist of
concentric shells where the singly charged particles form the inner
shells in order to minimize repulsion. Drocco {\em et al.} used
Langevin MD simulations for mixed clusters containing
$N_s=3$--34 singly charged and $N_d=5$, 6 and 7 doubly charged
particles, and they also determined the melting points. These authors
concluded that the melting behaviour of these binary trapped particles
strongly depends on whether the numbers of 
particles in the inner and outer shells are commensurate. 
They also found that the highest
melting points were obtained for $N_s=N_d+1$, where the corresponding
structures all have a single particle at the centre. 

\begin{figure}[htb]
\setlength{\epsfxsize}{9cm}
\centerline{\epsffile{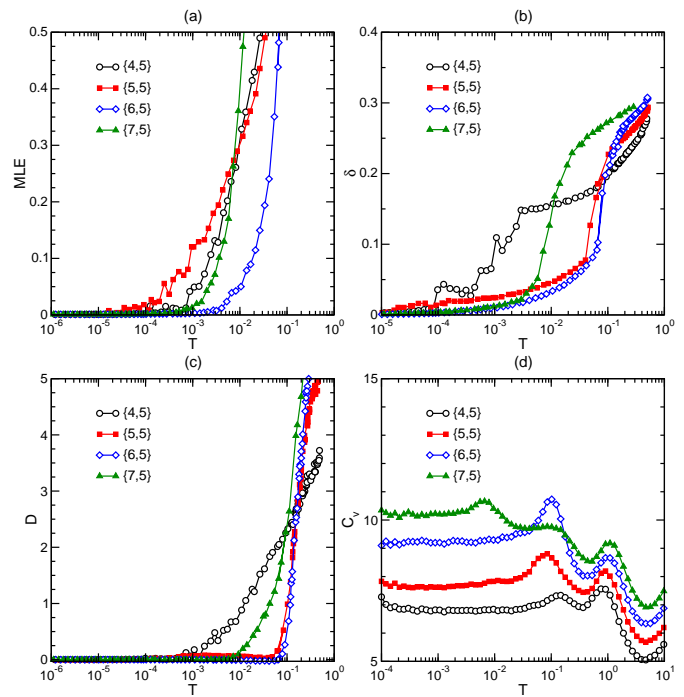}}
\caption{Dynamic and thermodynamic properties of some binary
\{$n$,5\} 2D Coulomb clusters with $n=4$--7 versus temperature.
(a) Maximum Lyapunov exponent (MLE) $\lambda$; (b) RMS bond length
fluctuation $\delta$; (c) diffusion coefficient $D$; (d) canonical
heat capacity $C_v$. The data in (a) to (c) are from microcanonical
MD simulations, the data in (d) are from canonical Monte Carlo simulations
and $C_v$ is in the units of Boltzmann constant.}
\label{fig:binary}
\end{figure}

We have performed MD simulations for these heterogeneous 
systems under similar conditions to those of the homogeneous clusters. In
Fig.~\ref{fig:binary}, $\lambda$, $D$, $\delta$ and the specific heat $C_v$
are plotted for \{4,5\}, \{5,5\}, \{6,5\} and \{7,5\}, selected as further
typical examples, highlighting differences in the stability and
dynamics. The onset of chaos occurs at temperatures close to those
seen in the homogeneous systems. \{6,5\} is the most stable cluster, in the
sense that it has the highest melting point in
Ref.~\onlinecite{drocco03}, while \{5,5\} is the most irregular.
Both the diffusion and Lindemann indices show the same qualitative
trends, namely that the melting points, as measured by the strong
increase in $\delta$, follow the  order $\{4,5\}<\{7,5\}<\{5,5\}<\{6,5\}$.

As for homogeneous clusters, the variation of the heat
capacities does not exactly match the dynamical
indices. Here the generic shape again starts at a constant value at low
temperature, which corresponds to harmonic behaviour. The high
temperature gas phase is still characterised by the same constant
value. In the intermediate temperature range, two main peaks are
associated with the melting of successive subclusters made of singly
charged, then doubly charged particles. Nearly one order of magnitude
separates the two associated melting temperatures. This interpretation
was further confirmed by calculating the heat capacity of the ternary
\{7,6,6\} cluster made of 7 singly charged particles, 6 doubly and 6
triply charged particles. For this cluster, three clear features
are observed in $C_v$, and only above $T\sim 2$ do the triply charged
particles mix with the rest of the cluster.

\begin{figure*}[htb]
\setlength{\epsfxsize}{14cm}
\centerline{\epsffile{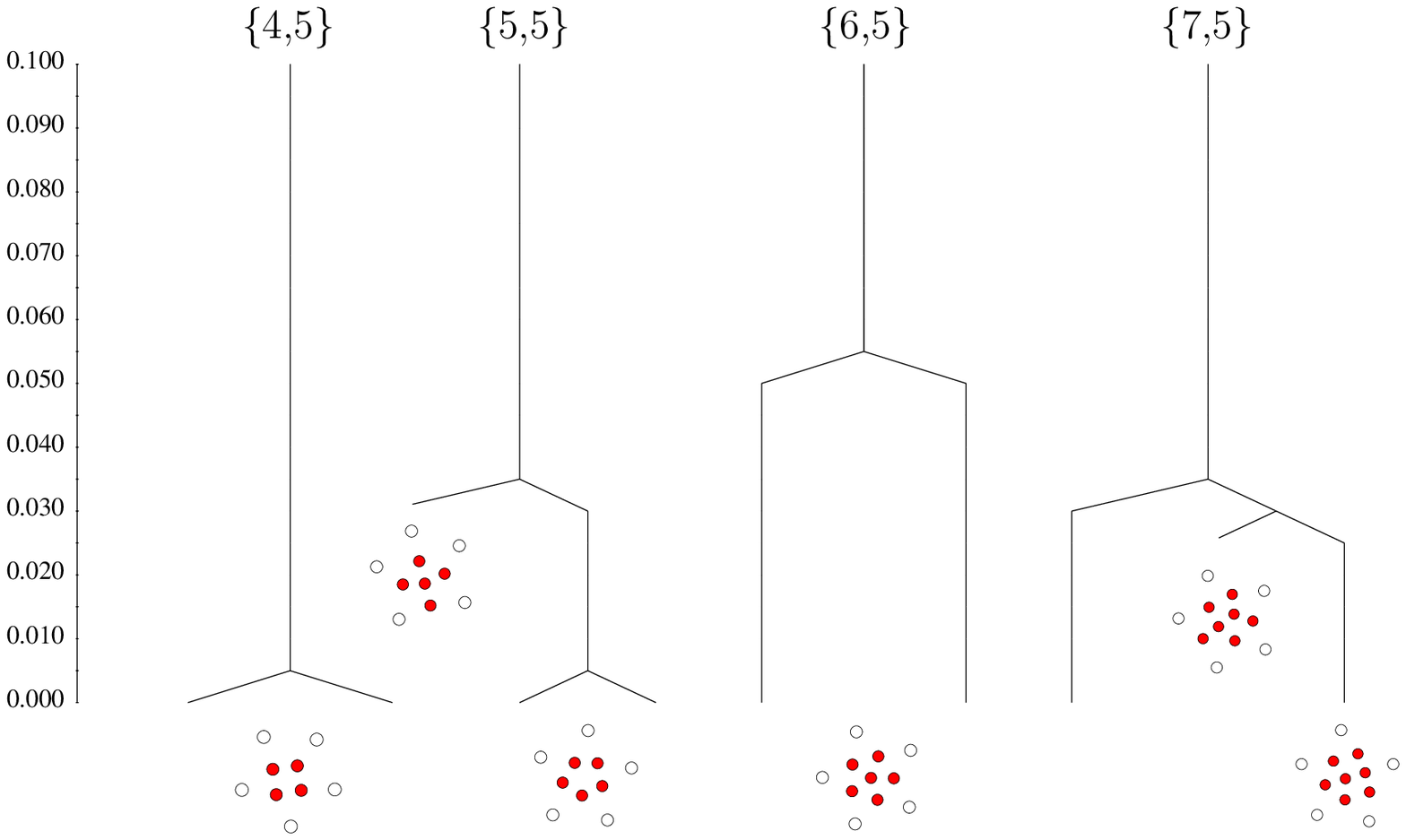}}
\caption{Disconnectivity graphs for the binary clusters of
Fig.~\protect\ref{fig:binary}. The global minimum energy is
shifted to zero for each size.}
\label{fig:treesb}
\end{figure*}

On top of this generic behaviour, a specific isomerization shoulder is
seen for the heat capacity of the \{7,5\} cluster in
Fig.~\ref{fig:binary}(d), at the same temperature $T\simeq 0.01$ where
the three dynamical parameters increase significantly. Again, the free
rotation of some internal shells revealed by $\lambda$, $D$ and
$\delta$ at low temperature has no thermal signature. The
disconnectivity graphs for these clusters are shown in
Fig.~\ref{fig:treesb}. As is particularly obvious from the graphs for
\{5,5\} and \{6,5\}, commensurate structure between the singly- and
doubly-charged shells is not the only factor needed to
explain the complicated behaviour exhibited by these finite Wigner
crystals. The extremely low barriers for the \{4,5\} and \{5,5\}
clusters permit facile rotation, which is correlated with the rapid increase
of all dynamical indicators at low temperatures. The \{6,5\} cluster is
especially resistant to isomerization because an extra, central
particle hinders the radial motion of the particles in the other shells.
Only for the \{7,5\} cluster is the isomerization energy low
enough to give rise to an extra peak in the heat capacity. This
case, and the case of the single-isomer \{6,5\} cluster, further show
that the multiple-peak structure of the heat capacity is not a
consequence of isomerizations. Therefore the general, two-peak shape of the
heat capacity is not related to isomerization, but is intrinsic
to the potential itself and the presence of two types of particles
organised in shells.

\begin{figure}[htb]
\setlength{\epsfxsize}{9cm}
\centerline{\epsffile{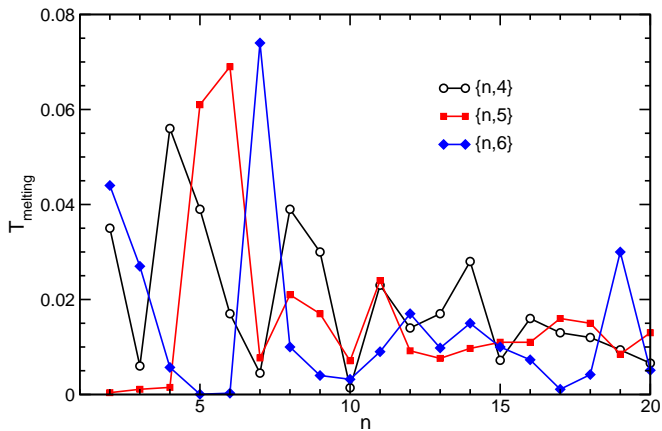}}
\caption{Size evolution of the melting point of the \{$n,4$\},
\{$n,5$\}, and \{$n,6$\} series of binary Coulomb clusters as obtained
from the Lindemann criterion.}
\label{fig:tmeltb}
\end{figure}

The variation of the melting temperature with cluster size is shown
in Fig.~\ref{fig:tmeltb} for the three \{$n$,4\}, \{$n$,5\} and \{$n$,6\}
binary series, with $2\leq n\leq 20$. Here the melting temperature was
defined according to the Lindemann criterion, as the temperature where
$\delta$ sharply increases above 0.1. The variation of the computed
melting point is qualitatively
similar to that reported by Drocco and coworkers
\cite{drocco03}. In particular, the high values for \{4,4\}, \{5,5\}, \{6,5\}
and \{7,6\} are reproduced. However, our results quantitatively
differ from those of Drocco {\em et al.} by a factor of at least 10.
Our data, obtained with exactly the same Hamiltonian as in
Ref.~\onlinecite{drocco03}, are in agreement with the computed
melting points for the homogeneous clusters, which are comparable with
previous results for the same systems
\cite{bedanov94}. Even though the present temperatures result from time
averages in our MD simulations, their range matches the canonical
temperature of our MC runs. Hence we believe that the data of
Fig.~\ref{fig:tmeltb} are reliable, and that the differences from the
results obtained by Drocco and coworkers are the consequence of an
unreported alternative choice of $A$ in the Hamiltonian used in
Ref.~\cite{drocco03}.

\begin{figure}[htb]
\setlength{\epsfxsize}{9cm}
\centerline{\epsffile{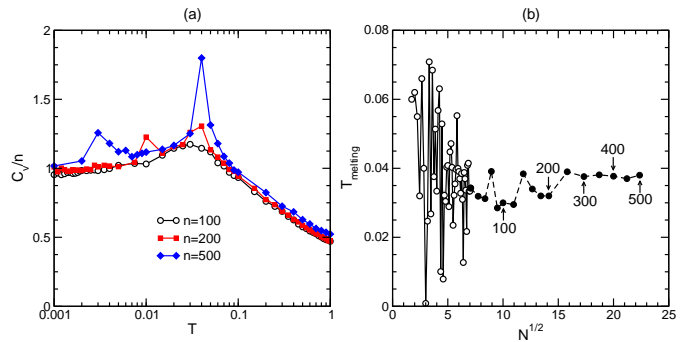}}
\caption{(a) Canonical heat capacities (in units of the Boltzmann constant) of some large homogeneous
clusters; (b) melting point of homogeneous clusters versus their radii.}
\label{fig:large}
\end{figure}

Lastly, we discuss the behaviour of larger clusters and how melting
evolves toward bulk. In Fig.~\ref{fig:large}(a) we 
show the heat capacities calculated from MC
simulations for the cluster sizes $n=100$, 200, and 500. The general
trend is that the main peak near $T\simeq 0.04$ becomes narrower and
sharper, suggesting that melting is a first-order
process rounded by finite  size effects \cite{labastie}.
However, the main heat capacity peak remains quite broad, so
a more detailed study should probably be performed on even larger
sizes, including a proper finite-size scaling analysis.
In addition,
extra features in the heat capacity are present at low temperatures,
even for the largest 500-particle cluster. Regular quenching confirms
that these premelting effects are due to isomerizations involving different
arrangements of the inner shells. At large sizes, even though the
global shape of the cluster is roughly circular, the radial ordering
into shells is replaced by a triangular close-packed lattice,
possibly with several defects \cite{lai99}. Rotation of the various
shells then becomes hindered, and the dynamics is reduced to more
usual consecutive isomerizations.

Fig.~\ref{fig:large}(a) supports the idea that a single, first-order
phase transition should occur in the bulk limit, if it could be defined. 
Since that there is no unified theory of 2D melting
\cite{ryzhov95}, we should probably be cautious in extrapolating the
heat capacities computed from Monte Carlo simulations. If we 
choose the peak of the heat capacity as the signature of melting, the
melting point thus defined shows very pronounced variations with the
cluster radius $n^{1/2}$, as illustrated in fig.~\ref{fig:large}(b).

The data for all sizes in the range $5\leq n\leq 50$ and for
larger clusters $(n=60,$ 70, 80, 90, 100, 120, 140, 160, 180, 200,
250, 300, 350, 400, 450, 500) are included in this graph. The strong
size dependence for small systems is similar to that observed for
the binary systems. However, in the range 50--500
the variation of the melting point seems less drastic than in the
smaller size range, even though numerous sizes are obviously missing
from the reported data. The bulk limit for the melting point is obtained to be 0.039.

This observation should not obscure the possibility that even large
sizes can exhibit a multiple stage melting. For example, significant
premelting peaks have been reported in large atomic clusters
\cite{calvo04}. The results obtained for $n=500$ show a similar
feature. Even though some premelting peaks are present in the caloric
curves of large clusters, a major peak seems to dominate above about
100 ions. This result may suggest that the transition between the
finite size regime and bulk \cite{cs2000} takes place near this size.

\section{Discussion and conclusion}

In this work, we have investigated some dynamic and thermodynamic
properties of finite 2D charged particles confined by a
parabolic trap. Our extensive data on homogeneous and heterogeneous
clusters were obtained from MD and MC simulations.
These data were interpreted by visualising the energy landscape through
disconnectivity graphs. Previous experimental and numerical findings
suggest that there are three types of motion for a particle: it can oscillate,
rotate in a relatively regular orbit, or make jumps between shells,
possibly causing isomerizations. These various phenomena were also
observed in the present work, but we found evidence that additional
`premelting' events due to structural transitions may also occur.
The various stages of melting create ambiguities in
defining the melting point, especially in small clusters.

Firstly, rotational motion is facile when very small energy
barriers separate the global energy minimum from itself, as is
apparent from the disconnectivity graphs for the 17- and 21-particle
homogeneous clusters, or the \{4,5\} binary cluster. Internal rotation
may give rise to significant jumps in the RMS bond length fluctuation,
sometimes above 0.1. Radial melting is then seen as yet another
increase in $\delta$ or other dynamical indicators at higher energy.
Therefore, the Lindemann criterion ($\delta \gtrsim 0.1$) 
is not fully reliable for detecting melting.

Secondly, rotational melting has no thermodynamic consequence since
only a single isomer is involved. The thermodynamics is sensitive to
the presence of different isomers, and also to the intrinsic shape of 
the potential felt by the particles. Irrespective of the isomers, the
heat capacity has a natural drop over a rather broad temperature range,
which marks the continuous transition toward the gas phase. Hence,
even a single cluster isomer can have heat capacities with complex
temperature dependence, as exemplified by the small \{4,5\}
binary clusters.

Thirdly, several isomers can give rise
to preliminary peaks in the heat capacity, which may sometimes obscure
the features associated with full radial melting. We found such a
case for the 31-particle cluster interacting through dipolar forces,
but very large clusters can also exhibit particularly strong
premelting peaks. The structural transition occurring in the
31-particle cluster has the same origins as the
fcc-icosahedral transition in the 3D 38-atom
Lennard-Jones cluster \cite{lj38wales,lj38neirotti}: the
second morphology has a much larger entropy than the global minimum. This
phenomenon has important consequences for the constant energy dynamical
behaviour, as this transition is not seen until sufficient energy
to surmount the isomerization barrier is present.
Thus there is a finite energy range where MD
simulations starting from the ground state structure are non-ergodic,
which is reflected by a strong shift in the estimated melting
temperature.

Most of our results illustrate how a simple and unique picture of
melting is hampered by finite size effects. Both the dynamical
parameters and the thermodynamical curves contribute to explaining
related but different aspects of melting in these 2D clusters. 
In particular, our work does not support a single parameter as a
universal measure for the melting point that could be used
unambiguously for all clusters. Nevertheless, the landscape approach
\cite{walesbook} that
has been followed here was found to provide useful insight for
interpreting and understanding the various numerical data.

The data gathered here for a broad range of sizes would seem to prevent a
general classification on the phenomenology of melting to be made. 
For example, the choice of molecular
dynamics or Monte Carlo method, and the
microcanonical or canonical ensemble, can change
the observed behavior significantly, particularly at small sizes where the
differences between the two statistical ensembles are the largest.
For instance, some structural transitions might not appear with
conventional MD methods, due to the disconnected character of phase space
at low total energies.

However, three main categories of behavior can be inferred from our results.
Especially in small clusters, a two-step orientational-then-radial
melting has been observed, with clear dynamical signatures, but only a
weak thermodynamical signature. Clusters exhibiting structural
transitions, on the other hand, may not involve any rotational melting
but display a strong change in their caloric curve. Finally, a more
conventional single-step melting is sometimes seen, as expected in
larger clusters, and also for smaller 'magic' sizes.

Our calculations indicate that significant finite size effects remain
in the thermodynamics of the largest clusters considered. 
However, above a few hundred 
ions, melting seems to be essentially a one-step process.
This result is consistent with the expectation that the dynamics becomes
more and more dominated by localised processes, due to the less
favourable rotational motion. This observation seems to justify the
use of the Lindemann index or other dynamical parameters for larger
systems. Additionally, the main heat capacity peak becomes larger, and
its maximum value coincides with the onset of disorder in the
dynamical indicators. Even though premelting features may remain
noticeable, we expect the onset of radial melting to become closer and
closer to orientational melting, and eventually similar to a bulk
first-order melting process.

\section {Acknowledgements}

We are grateful for a collaborative research grant given by CNRS and TUBITAK.

\end{document}